\documentclass[twocolumn,showpacs,floatfix]{revtex4}
\usepackage{graphics}

\begin{document}

\title{Quantum properties of counter-propagating two-photon states generated
in a planar waveguide.}

\author{Jan Pe\v{r}ina Jr. \thanks{e-mail:
perinaj@prfnw.upol.cz} \\
Joint Laboratory of Optics
of Palack\'{y} University and \\
Institute of Physics of Academy of Sciences of the Czech
Republic, \\
17. listopadu 50A, 772 00 Olomouc,
Czech Republic.}

\begin{abstract}
A nonlinear planar waveguide pumped by a beam orthogonal to its
surface may serve as a versatile source of photon pairs. Changing
pump-pulse duration, pump-beam transverse width, and angular
decomposition of pump-beam frequencies characteristics of a photon
pair including spectral widths of signal and idler fields, their
time durations as well as degree of entanglement of two fields can
be changed significantly. Using the measured spectral widths of
the down-converted fields and width of a coincidence-count dip in
a Hong-Ou-Mandel interferometer entropy of entanglement can be
determined.
\end{abstract}



\maketitle

\section{Introduction}

The process of spontaneous parametric down-conversion as a source
of entangled photon pairs has been used in numerous experiments
during the last twenty years. The physicists went through a long
way from the first pioneering experiments showing basic properties
of entangled photon pairs [spectral (temporal) and polarization
correlations of photons comprising a pair, see, e.g.
\cite{Hong1987,Mandel1995}] to the recent sophisticated
experimental setups that demonstrate quantum teleportation
\cite{Bouwmeester1997}, quantum cloning \cite{DeMartini2000}, test
Bell and other nonclassical inequalities
\cite{Perina1994,Bovino2005}, or generate
Greenberger-Horne-Zeilinger states \cite{Bouwmeester1999}.
Usefulness of the fragile entangled photon pairs has also been
verified in applications like quantum cryptography
\cite{Bruss2000} and absolute metrology \cite{Migdal1999} to name
a few.

Also sources of photon pairs have been improved significantly.
Various geometric configurations of usual nonlinear crystals (see,
e.g., in \cite{Kwiat1999,Nambu2002}) are gradually being replaced
by new more efficient sources based on quasi-phase-matching
\cite{Kuklewicz2005,Carrasco2004,Chatellus2006,Harris2007}.
Nonlinear photonic-band-gap fibers seem to be extraordinarily
interesting as sources of photon pairs (emerging in the process of
four-wave mixing) due to a high effective nonlinearity
\cite{Li2005,Fulconis2005}. Nonlinear layered structures as
sources of photon pairs are under investigation at present
\cite{Centini2005,PerinaJr2006}.

A great deal of attention has been devoted to the generation of
two-photon states with specific spectral properties in bulk
materials. Entangled two-photon states with coincident frequencies
have been obtained using extended phase-matching conditions, i.e.
assuming group-velocity phase matching on the top of the usual
phase matching \cite{Giovannetti2002,Giovannetti2002a,Kuzucu2005}.
Such states have perfect visibility in the usual interferometric
setups and moreover allow a very precise clock synchronization
\cite{Giovannetti2001}. Spectrally uncorrelated two-photon states
have also been studied extensively because they seem to be
extraordinarily useful for linear quantum computation that needs
indistinguishable photons with a perfect time synchronization
\cite{URen2003,URen2005}. Although such states can be generated
from usual bulk crystals for suitable crystal length and pump-beam
waist in non-collinear configurations
\cite{Grice2001,Carrasco2006}, more flexible approaches have been
suggested in \cite{URen2003} exploiting phase-matching in the
transverse plane. Both coincident-frequency entangled and
unentangled two-photon states can be obtained in a nonlinear
crystal with achromatic phase matching. i.e. when pump-beam
frequencies are decomposed such that every frequency propagates
along a slightly different angle \cite{Torres2005,Torres2005a}.
Also the so-called nonlinear crystal superlattices, i.e.
structures composed of several identical pieces of nonlinear
material and spacers, have provided additional degrees of freedom
for tailoring properties of the generated two-photon states
\cite{URen2005a,URen2006}. Results appropriate for photonic-band
gap structures \cite{PerinaJr2007} are reached for a higher number
of nonlinear pieces.

Also planar nonlinear structures are promising as sources of
entangled photon pairs because, using specific geometric
configurations, they provide large possibilities for tailoring
properties of the generated photon pairs. One of the most
perspective configurations (suggested and elaborated in
\cite{Ding1995,DeRossi2002,Booth2002,Walton2003}) is based upon
pumping a planar waveguide by a beam perpendicular to its surface.
Signal and idler photons then emerge as counter-propagating guided
waves. Disadvantage of this configuration is that the pump beam
propagates through a very thin nonlinear medium which thickness is
given by the depth of the waveguide. In order to suppress
destructive interference in the three-wave nonlinear process this
thickness has to be of the order of pump-field wavelength. Thus,
very low generation rates have to be expected. To cope with a low
efficiency of the nonlinear process more sophisticated structures
have been suggested \cite{Ravaro2005,Sciscione2006}. They use
Bragg mirrors both above and below the waveguide. Their properties
are chosen such that the pump-beam electric-field amplitude is
maximally enhanced inside the nonlinear waveguide. This may result
in enhancement of the efficiency by several orders of magnitude.
The first experimental demonstration of this source has been
already reported in \cite{Lanco2006}. The use of pump pulses with
frequencies propagating along different angles brings even more
flexibility and so photon pairs with an arbitrary shape of a
two-photon spectral amplitude can be generated \cite{Walton2004}.

As shown in this paper using a simple model of planar waveguide
with parabolic index of refraction \cite{Snyder1983,Booth2002},
properties of photon pairs (spectral and temporal widths of the
down-converted fields and entanglement) generated from this type
of geometry can be modified in broad ranges simply by changing
parameters of the pump beam (pump-pulse duration, pump-beam
transverse width, angular decomposition of pump-beam frequencies).
Spectral widths of the down-converted fields can spread from circa
1 nm up to several tens of nm. Spectrally uncorrelated (separable)
states as well as strongly entangled states can be observed. A
method for the determination of entropy of entanglement from the
measured signal- and idler-field intensity spectra and width of
the coincidence-count pattern in a Hong-Ou-Mandel interferometer
is also suggested.

The paper is organized as follows. Sec. II is devoted to the
determination of a two-photon spectral amplitude of the generated
photon pair. This amplitude is later used to derive spectral (Sec.
III) and temporal (Sec. IV) properties of the down-converted
fields, and characterize entanglement of the signal and idler
fields (Sec. V). Experimental determination of entropy of
entanglement is discussed in Sec. VI. Conclusions are drawn in
Sec. VII. Appendix A contains general formulas describing
properties of photon pairs. Appendix B is devoted to Schmidt
decomposition of a two-photon spectral amplitude.

\section{Two-photon spectral amplitude of a photon pair generated from a
waveguide}

\begin{figure}    
 \resizebox{0.6\hsize}{!}{\includegraphics{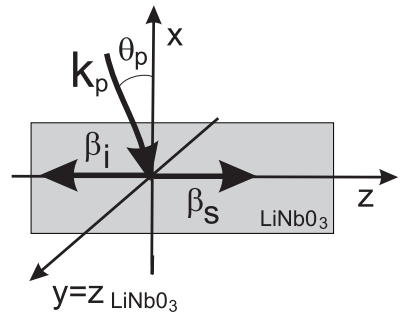}}
 \caption{Sketch of a nonlinear planar waveguide made of LiNbO$
 {}_3 $ with parabolic profile of index of refraction along the $ x $
 axis. The optical axis of LiNbO$ {}_3 $ as well as polarization directions
 of signal, idler, and pump fields are parallel to the $ y $ axis.
 The signal (idler) field propagates along the $ +z $ ($ -z $) axis
 with propagation constant $ \beta_s $ ($ \beta_i $); the pump field
 with central wave-vector $ k_p^0 $ propagates under central angle
 $ \theta_p^0 $
 with respect to the $ x $ axis. The waveguide is confined into the
 region $ (-L_y/2,L_y/2) $ in $ y $ direction.}
\label{fig1}
\end{figure}

We consider the generation of a photon-pair into guided modes of a
planar waveguide made of LiNbO$ {}_3 $ that is pumped by a
travelling-wave pump beam at the wavelength $ \lambda_p = 1.064
\times 10^{-6} $~m that propagates under the central angle $
\theta_p^0 $ with respect to the $ x $ axis orthogonal to the
surface (see Fig.~\ref{fig1}). The pump beam is assumed not to be
cross spectrally pure in general, i.e. different pump-beam
frequencies can propagate under different propagation angles. The
guided signal and idler fields then form counter-propagating
beams.
For simplicity, we pay attention to the waveguide with a parabolic
profile of index of refraction $ n(x) $, $ n(x)^2 = n_0^2
(1-\alpha^2 x^2) $ ($ \alpha $ is parameter of the waveguide) that
supports only TE-guided modes with a gaussian profile.

Energy of the nonlinear interaction that produces photon pairs is
described by Hamiltonian $ \hat{H} $:
\begin{equation}   
 \hat{H}(t) = \epsilon_0 d \int dV \left[ E^{(+)}_p({\bf r},t)
 \hat{E}^{(-)}_s({\bf r},t) \hat{E}^{(-)}_i({\bf r},t) +
  {\rm h.c.} \right] ,
\label{1}
\end{equation}
where $ E^{(+)}_p $ is positive-frequency part of the pump-beam
electric-field amplitude and $ \hat{E}^{(-)}_s $ ($
\hat{E}^{(-)}_i $) stands for negative-frequency part of the
signal- (idler-) beam electric-field amplitude operator. Symbol $
\epsilon_0 $ denotes permittivity of vacuum, $ d $ is effective
second-order nonlinear coefficient, and $ {\rm h.c.} $ means a
hermitian conjugated term. Integration in Eq.~(\ref{1}) is over
interaction volume $ V $.

In the considered waveguide, positive-frequency parts of the
signal- and idler-beam electric-field amplitude operators $
\hat{E}^{(+)}_s $ and $ \hat{E}^{(+)}_i $ can be decomposed as [$
E_p^{(-)} = (E_p^{(+)})^\dagger $]:
\begin{eqnarray}   
 \hat{E}^{(+)}_a({\bf r},t) &=& \int d\omega_a e_a({\bf r},\omega_a)
  \hat{a}_a(\omega_a) , \nonumber \\
 e_a({\bf r},\omega_a) &=& C_a(\omega_a) {\rm rect}_{(-L_y/2,L_y/2)}(y)
  \nonumber \\
 & & \hspace{-5mm} \mbox{} \times \exp\left(-\frac{\gamma_a^2x^2}{2}
  \right) \exp(\pm i\beta_a z)
  \exp(-i\omega_a t) , \nonumber \\
 |C_a(\omega_a)|^2 &=& \frac{\hbar\omega_a \gamma_a}{2\sqrt{\pi}
  \epsilon_0 n_0^3(\omega_a) c L_y}, \hspace{5mm} a=s,i ;
\label{2}
\end{eqnarray}
symbol $ \hat{a}_a(\omega_a) $ denotes annihilation operator of a
mode with frequency $ \omega_a $ in field $ a $. The sign $ + $ ($
- $) in the second relation in Eq.~(\ref{2}) is for the signal
(idler) field that propagates along the $ +z $ ($ -z $) axis. In
Eq.~(\ref{2}), $ n_0(\omega_a) $ is index of refraction of field $
a $ with frequency $ \omega_a $, $ c $ means speed of light in
vacuum, $ \hbar $ reduced Planck constant, $ L_y $ width of the
waveguide along the $ y $ axis where a rectangular profile is
assumed, and $ \gamma_a(\omega_a) = \sqrt{n_0(\omega_a) \omega_a
\alpha /c} $. Function $ {\rm rect}_{(a,b)}(x) $ equals 1 for $ a
< x < b $ and is zero otherwise. Normalization constant $ C_a $ in
Eq.~(\ref{2}) has been determined from the condition that a photon
emitted into a guided mode has energy $ \hbar\omega_a $:
\begin{equation}   
 2 \epsilon_0 n_0^2(\omega_a) \int dV | e_a({\bf r},\omega_a)|^2
  = \hbar \omega_a , \hspace{5mm} a=s,i .
\end{equation}
We assume that the interaction volume $ V $ has `length' $ c /
n_0(\omega_a^0) $ along the $ z $ axis when determining constant $
C_a $ in Eq.~(\ref{2}).

Propagation constant $ \beta_a $ of field $ a $ along the $ z $
axis is given as follows (for details, see \cite{Snyder1983}):
\begin{equation}    
 \beta_a(\omega_a) = \frac{n_0(\omega_a)\omega_a}{c}
 \sqrt{ 1 - \frac{\alpha c}{n_0(\omega_a)\omega_a} }.
\label{4}
\end{equation}
It can be approximately expressed as:
\begin{eqnarray}    
 \beta_a(\omega_a) &=& \beta_a^0 +
  \frac{\omega_a-\omega_a^0}{v_a}, \nonumber \\
 \beta_a^0 &=& \beta_a(\omega_a^0) , \nonumber \\
 \frac{1}{v_a} &=& \left. \frac{d\beta_a}{d\omega_a}
 \right|_{\omega_a = \omega_a^0},
\end{eqnarray}
where $ v_a $ denotes group velocity of field $ a $.

The travelling-wave pump beam with central frequency $ \omega_p^0
$ and propagating along central angle $ \theta_p^0 $ is assumed to
have a gaussian profile along the $ y $ and $ z $ axes
characterized by widths $ Y_p $ and $ Z_p $, respectively, and is
also gaussian in the time domain with pump-pulse duration $ \tau_p
$ and chirp parameter $ a_p $. Different monochromatic components
of the pump beam can propagate along different angles $
\theta_p(\omega_p) $, e.g. as a consequence of pump-beam
reflection on an optical grating or after propagation through a
prism. The pump-beam positive-frequency electric-field amplitude $
E_p^{(+)} $ can be written in the form:
\begin{eqnarray}  
 E_p^{(+)}({\bf r},t) &=& \frac{1}{\sqrt{2\pi}} \int d\omega_p
  E_p^{(+)}({\bf r},\omega_p) \exp(-i\omega_p t) , \nonumber \\
 E_p^{(+)}({\bf r},\omega_p) &=& C_p(\omega_p) \exp\left( -\frac{z^2}{Z_p^2}
  \right) \exp\left( -\frac{y^2}{Y_p^2} \right) \nonumber \\
 & & \mbox{} \hspace{-10mm} \times
  \exp[ik_p\sin(\theta_p(\omega_p))z] \exp[-ik_p\cos(\theta_p(\omega_p))x]
  \nonumber \\
 & & \mbox{} \hspace{-10mm} \times
  \exp\left[- \frac{\tau_p^2(\omega_p-\omega_p^0)^2}{4(1+ia_p)}
  \right], \nonumber \\
 |C_p(\omega_p)|^2 &=& \nonumber \\
  & & \hspace{-10mm} \frac{\tau_p}{\sqrt{2\pi}\pi \epsilon_0
  n_0^2(\omega_p)Y_p Z_p(1+a_p^2)} \frac{P_p}{v_p\cos(\theta_p(\omega_p)) f};
  \nonumber \\
 & &
\label{6}
\end{eqnarray}
$ v_p $ is pump-field group velocity [$ k_p(\omega_p) = k_p^0 +
(\omega_p-\omega_p^0)/v_p $, $ k_p^0 = k_p(\omega_p^0) $, $ 1/v_p
= dk_p/d\omega_p|_{\omega_p=\omega_p^0} $], $ P_p $ pump-field
power, and $ f $ denotes repetition rate of the pulsed pump field.

First-order perturbation solution of the Schr\"{o}dinger equation
using Hamiltonian $ H(t) $ given in Eq.~(\ref{1}) and assuming the
incident signal and idler fields in vacuum states provides the
following expression for an outgoing two-photon state $
|\psi^{(2)} \rangle $:
\begin{eqnarray}    
 |\psi^{(2)} \rangle = \int d\omega_s \int d \omega_i \,
 \Phi^{1p}(\omega_s,\omega_i) \hat{a}_s^\dagger(\omega_s)
  \hat{a}_i^\dagger(\omega_i) |{\rm vac}\rangle .
\end{eqnarray}
Two-photon spectral amplitude $ \Phi^{1p}(\omega_s,\omega_i) $
giving the probability amplitude of having a signal photon at
frequency $ \omega_s $ and an idler photon at frequency $ \omega_i
$ generated from one pump pulse is derived in the form:
\begin{eqnarray}   
 \Phi^{1p}(\omega_s,\omega_i) &=& -i \frac{\sqrt{2\pi}2\pi^2 \epsilon_0 d}{\hbar}
  C_p(\omega_s+\omega_i) C_s^*(\omega_s) C_i^*(\omega_i)
  \nonumber \\
 & & \hspace{-15mm} \mbox{} \times
  \frac{Y_p Z_p}{\sqrt{\gamma_s^2+\gamma_i^2}} \,
  {\rm erf}\left(\frac{L_y}{2Y_p}\right) \exp\left[-\frac{\tau_p^2
  (\omega_s+\omega_i-\omega_p^0)^2}{4(1+ia_p)}\right] \nonumber \\
 & & \hspace{-15mm} \mbox{} \times \exp\left[-\frac{Z_p^2[k_p\sin(\theta_p(\omega_s+\omega_i))
  -\beta_s+\beta_i]^2}{4}\right] \nonumber \\
 & & \hspace{-15mm} \mbox{} \times
  \exp\left[-\frac{k_p^2\cos^2(\theta_p(\omega_s+\omega_i))}{2(\gamma_s^2+\gamma_i^2)}
  \right] ;
\end{eqnarray}
symbol $ {\rm erf} $ stands for error function [$ {\rm erf}(x) =
2/\sqrt{\pi} \int_{0}^{x} \exp(-y^2) dy $].

Using first-order Taylor expansions for propagation constants $
k_p $, $ \beta_s $, and $ \beta_i $ as well as for propagation
angle $ \theta_p $ together with second-order Taylor expansion for
the expression $ 1/(\gamma_s^2 + \gamma_i^2) $ [ $ 1/(\gamma_s^2 +
\gamma_i^2) \approx g_0 + g_{1s}\Delta\omega_s +
g_{1i}\Delta\omega_i + g_{2s}\Delta\omega_s^2 +
g_{2i}\Delta\omega_i^2 + g_{2si}\Delta\omega_s\Delta\omega_i $, $
\Delta\omega_a = \omega_a-\omega_a^0 $, $ a=s,i $] we arrive at a
two-photon spectral amplitude $ \Phi $ giving contribution from $
f $ pump pulses and having the following gaussian form:
\begin{eqnarray}   
 \Phi(\omega_s,\omega_i) &=& C_\Phi \sqrt{\frac{Z_p\tau_p}{1+a_p^2}}
  \exp[-\phi(\omega_s,\omega_i)]
  , \nonumber \\
 \phi(\omega_s,\omega_i) &=& f_{2s}\Delta\omega_s^2 + f_{2i}\Delta\omega_i^2
  + f_{2si}\Delta\omega_s\Delta\omega_i \nonumber \\
 & & \mbox{} + f_{1s}\Delta\omega_s
  + f_{1i}\Delta\omega_i + f_0.
\label{TPAO}
\end{eqnarray}
Coefficients $ f $ occurring in Eq.~(\ref{TPAO}) are expressed as
follows:
\begin{eqnarray}   
 f_{2a} &=& \frac{\tau_p^2}{4(1+ia_p)} + \frac{V_{pa}^2 Z_p^2}{4} +
 \frac{1}{\sigma_a^2} + {\cal G}_a , \nonumber \\
 & & \hspace{-7mm} {\cal G}_a = \frac{k_p^0 \cos(\theta_p^0)}{2}
  \Biggl( k_p^0 \cos(\theta_p^0) g_{2a} \nonumber \\
  & & \mbox{} + 2 \left[ \frac{\cos(\theta_p^0)}{v_p} -
   k_p^0\sin(\theta_p^0)\tilde{D}_{\theta_p}
  \right] g_{1a} \nonumber \\
  & & \mbox{} +  \left[ \frac{\cos(\theta_p^0)}{k_p^0 v_p^2} -
   \frac{4\sin(\theta_p^0)}{v_p} \tilde{D}_{\theta_p} \right. \nonumber \\
  & & \left. \mbox{} -
   \frac{ k_p^0 \cos(2\theta_p^0) }{ \cos(\theta_p^0)} \tilde{D}_{\theta_p}^2
   \right] g_0 \Biggr) , \nonumber \\
  & & \hspace{4cm} a=s,i, \nonumber \\
 f_{2si} &=& \frac{\tau_p^2}{2(1+ia_p)} + \frac{V_{ps}V_{pi} Z_p^2}{2} +
   \frac{1}{\sigma_a^2} + {\cal G}_{si} , \nonumber \\
  & & \hspace{-7mm} {\cal G}_{si} = \frac{k_p^0 \cos(\theta_p^0)}{2}
   \Biggl( k_p^0 \cos(\theta_p^0) g_{2si} \nonumber \\
   & & \mbox{} +
    2 \left[ \frac{\cos(\theta_p^0)}{v_p} - k_p^0\sin(\theta_p^0)\tilde{D}_{\theta_p} \right]
    (g_{1s}+g_{1i})  \nonumber \\
   & & \mbox{} + 2 \left[ \frac{\cos(\theta_p^0)}{k_p^0 v_p^2} - \frac{4\sin(\theta_p^0)}{v_p}
    \tilde{D}_{\theta_p} \right. \nonumber \\
   & & \left. \mbox{}  - \frac{k_p^0 \cos(2\theta_p^0)}{ \cos(\theta_p^0)} \tilde{D}_{\theta_p}^2
    \right] g_0 \Biggr) , \nonumber \\
   & & \hspace{4cm} a=s,i, \nonumber \\
 f_{1a} &=& k_p^0\cos(\theta_p^0) \Biggl(
  \frac{k_p^0\cos(\theta_p^0)}{2} g_{1a} \nonumber \\
  & & \mbox{} \hspace{-5mm}  + \left[ \frac{\cos(\theta_p^0)}{v_p} - k_p^0\sin(\theta_p^0) \tilde{D}_{\theta_p}
  \right] g_0 \Biggr)  , \hspace{5mm} a=s,i,
  \nonumber \\
 f_0 &=& \frac{[k_p^0\cos(\theta_p^0)]^2}{2} g_0 ,
\label{10}
\end{eqnarray}
and
\begin{eqnarray}   
 V_{ps} = \frac{\sin(\theta_p^0)}{v_p} + k_p^0 \cos(\theta_p^0) \tilde{D}_{\theta_p}
  - \frac{1}{v_s} , \nonumber
  \\
 V_{pi} = \frac{\sin(\theta_p^0)}{v_p} + k_p^0 \cos(\theta_p^0) \tilde{D}_{\theta_p}
  + \frac{1}{v_i}.
\label{11}
\end{eqnarray}
Coefficient $ \tilde{D}_{\theta_p} $ describes angular
decomposition of pump-beam frequencies; $ \tilde{D}_{\theta_p} =
\left. d\theta_p(\omega_p)/d\omega_p \right|_{\omega_p=\omega_p^0}
$; $ D_{\theta_p} = \tilde{D}_{\theta_p}(\omega_p^0)^2 / (2\pi c)
$. Influence of frequency filters with a gaussian shape is
described by their widths $ \sigma_s $ and $ \sigma_i $ (for the
signal and idler fields, respectively) that occur in
Eq.~(\ref{10}).

Normalization constant $ C_\Phi $ introduced in Eq.~(\ref{TPAO})
is determined along the expression:
\begin{eqnarray}   
 |C_\phi|^2 &=& \frac{ \sqrt{2\pi}\pi^2 d^2\omega_s^0\omega_i^0 }{\epsilon_0 c^2
  n_0^2(\omega_p^0) n_0^3(\omega_s^0) n_0^3(\omega_i^0)}
  \frac{\sqrt{n_0(\omega_s^0)n_0(\omega_i^0)\omega_s^0\omega_i^0}}{
   n_0(\omega_s^0)\omega_s^0 + n_0(\omega_i^0)\omega_i^0}
   \nonumber \\
  & & \mbox{} \times
  \frac{Y_p}{L_y^2}{\rm erf}^2\left( \frac{L_y}{2Y_p} \right)
   \frac{P_p}{v_p\cos(\theta_p^0)} .
\label{12}
\end{eqnarray}

Phase matching for central frequencies has been assumed when
deriving the expression for two-photon spectral amplitude $ \Phi $
written in Eq.~(\ref{TPAO}), i.e.
\begin{equation}   
 k_p^0 \sin(\theta_p^0) - \beta_s^0 + \beta_i^0 = 0 .
\label{PM}
\end{equation}
Equation (\ref{PM}) represents condition for possible values of
central frequencies $ \omega_p^0 $, $ \omega_s^0 $, $ \omega_i^0 $
and central angle $ \theta_p^0 $ of pump-beam propagation. This
condition even with the inclusion of quasi-phase matching has been
extensively studied in \cite{Booth2002}.

The role of pump-pulse duration $ \tau_p $ and pump-beam
transverse width $ Z_p $ on the shape of two-photon spectral
amplitude $ \Phi $ can be understood when we transform the
amplitude $ \Phi $ into new variables $ \Omega $ and $ \omega $; $
\Omega = (\omega_s + \omega_i)/2 $, $ \omega = (\omega_s -
\omega_i)/2 $:
\begin{eqnarray}   
 \Phi(\Omega,\omega) &=& 2C_\phi \sqrt{\frac{Z_p\tau_p}{1+a_p^2}} \nonumber \\
 & & \hspace{-10mm} \mbox{} \times  \exp \Biggl\{ - \left[
  \frac{\tau_p^2}{1+ia_p} + \frac{Z_p^2 ( V_{ps} + V_{pi})^2}{4} +
  \frac{1}{\sigma_+^2}
  \right] \Delta\Omega^2  \nonumber \\
 & & \hspace{-10mm} \mbox{} + \left[ \frac{Z_p^2 (V_{ps} + V_{pi})V_{si}}{2}
  - \frac{2}{\sigma_-^2} \right] \Delta\Omega \Delta\omega  \nonumber \\
 & & \hspace{-10mm} \mbox{} -  \left[ \frac{Z_p^2 V_{si}^2}{4} + \frac{1}{\sigma_+^2} \right]
  \Delta\omega^2  \Biggr\},
\label{TPAOo}
\end{eqnarray}
where
\begin{eqnarray}   
 V_{si} &=& \frac{1}{v_s} + \frac{1}{v_i} , \\
 \frac{1}{\sigma_+^2} &=& \frac{1}{\sigma_s^2} +
  \frac{1}{\sigma_i^2}, \nonumber \\
 \frac{1}{\sigma_-^2} &=& \frac{1}{\sigma_s^2} -
  \frac{1}{\sigma_i^2} .
\end{eqnarray}
Coefficients $ {\cal G}_s $, $ {\cal G}_i $, and $ {\cal G}_{si} $
occurring in Eq.~(\ref{10}) have been neglected when the
expression in Eq.~(\ref{TPAOo}) has been derived because they are
small in comparison with those written explicitly in
Eq.~(\ref{TPAOo}) under the studied conditions. We can see from
Eq.~(\ref{TPAOo}) that pump-pulse duration $ \tau_p $ influences
only the coefficient of quadratic form in sum frequency $ \Omega
$, whereas pump-beam transverse width $ Z_p $ and widths of
spectral filters $ \sigma_s $ and $ \sigma_i $ modify all of them.
The shape of a two-photon amplitude $ \Phi $ can also be
controlled using parameter $ \tilde{D}_{\theta_p} $ of angular
decomposition of pump-beam frequencies that occurs in expressions
for the coefficients multiplying $ \Delta\Omega^2 $ and $
\Delta\Omega\Delta\omega $ in Eq.~(\ref{TPAOo}). Nonzero values of
parameter $ \tilde{D}_{\theta_p} $ lead to rotation of the shape
of the two-photon amplitude $ \Phi $ in the plane spanned by
frequencies $ \omega_s $ and $ \omega_i $
\cite{Torres2005,Torres2005a}.

Considering frequency degenerate case ($ \omega_s^0 = \omega_i^0
$, i.e. $ v_s = v_i $ and $ \theta_p^0 = 0 $), cross spectrally
pure pump beam ($ \tilde{D}_{\theta_p} = 0 $), and  omitting
frequency filters ($ \sigma_s, \sigma_i \rightarrow \infty $) we
obtain the two-photon spectral amplitude $ \Phi $ in a simple
form:
\begin{eqnarray}      
 \Phi(\Omega,\omega) &=& 2C_\phi \sqrt{\frac{Z_p\tau_p}{1+a_p^2}} \nonumber \\
 & & \mbox{} \hspace{-3mm} \times \exp \left[ -
  \frac{\tau_p^2}{1+ia_p} \Delta\Omega^2 - \frac{Z_p^2 V_{si}^2}{4}
  \Delta\omega^2  \right];
\end{eqnarray}
i.e. pump-pulse duration $ \tau_p $ determines properties
depending on sum frequency $ \Omega $ whereas pump-beam transverse
width $ Z_p $ is responsible for properties related to difference
frequency $ \omega $. This behavior is similar to that occurring
in coincident-frequency entangled two-photon states generated from
bulk materials and described, e.g., in \cite{Giovannetti2002}
(length $ L $ of a crystal plays the role of $ Z_p $).

\section{Spectral properties of photon pairs, photon-pair generation rate}

\begin{figure}    
 \resizebox{0.9\hsize}{!}{\includegraphics{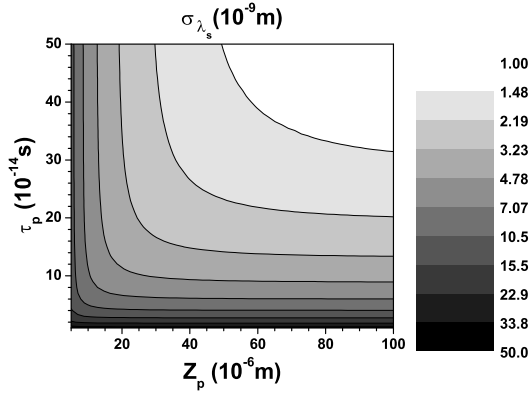}}
 \caption{Contour plot of width $ \sigma_{\lambda_s} $ of the signal-field
  intensity spectrum as a function of pump-pulse duration $ \tau_p $ and
  pump-beam transverse width $ Z_p $ is shown; logarithmic scale is used
  on the $ z $ axis; $ \lambda_p^0 = 1.064
  \times 10^{-6} $~m, $ \lambda_s^0=\lambda_i^0 = 0.532
  \times 10^{-6} $~m, $ P_p = 1 $~W, $
  a_p = 0 $, $ Y_p = 1\times 10^{-5} $~m, $ L_y = 1\times 10^{-5} $~m,
  $ D_{\theta_p} = 0 $~deg m$ {}^{-1} $,
  $\alpha = 4\times 10^6 $~m$ {}^{-1} $, $\sigma_s = \sigma_i \rightarrow
  \infty $, $ d = 41.05 \times 10^{-12} $~mV$ {}^{-1} $.}
\label{fig2}
\end{figure}

Number $ N $ of photon pairs generated in 1~s is determined along
the formula:
\begin{equation}  
 N = \int_{-\infty}^{\infty} d \omega_s \int_{-\infty}^{\infty} d
 \omega_i |\Phi(\omega_s,\omega_i)|^2.
\label{N}
\end{equation}
The general expression for the number $ N $ of photon pairs
assuming the spectral two-photon amplitude $ \Phi $ in the form
written in Eq.~(\ref{TPAO}) can be found in Appendix A
[Eq.~(\ref{A1})]. Neglecting coefficients $ {\cal G}_s $, $ {\cal
G}_i $, and $ {\cal G}_{si} $ in Eq.~(\ref{10}), a simplified
expression can be derived:
\begin{equation}   
 N = |C_\Phi|^2 \frac{\pi Z_p\tau_p}{(1+a_p^2)\sqrt{D_{f^r}}},
\label{19}
\end{equation}
where
\begin{eqnarray}  
 D_{f_r} &=& \frac{4}{\sigma_s^2\sigma_i^2} +
  \frac{\tau_p^2}{(1+a_p^2)}\left(\frac{1}{\sigma_s^2}+\frac{1}{\sigma_i^2}\right)
  \nonumber \\
 & & \mbox{} +
  \frac{\tau_p Z_p^2 V_{si}^2}{4(1+a_p^2)} +
  \frac{Z_p^2 V_{pi}^2}{\sigma_s^2} + \frac{Z_p^2
  V_{ps}^2}{\sigma_i^2} .
\label{20}
\end{eqnarray}
If frequency filters are omitted, the expression in Eq.~(\ref{19})
for photon-pair generation rate $ N $ gets a simple form:
\begin{equation}   
 N = |C_\Phi|^2 \frac{2\pi^2}{\sqrt{1+a_p^2}V_{si}},
\end{equation}
i.e. the generation rate $ N $ does not depend both on pump-pulse
duration $ \tau_p $ and pump-beam transverse width $ Z_p $. This
is a consequence of geometry of the considered three-mode
interaction. Assuming values of the waveguide parameters as
written in Fig.~\ref{fig2} and incident pump-field power $ P_p = 1
$~W, the generation rate $ N $ equals $ 3 \times 10^4 $~s$ {}^{-1}
$, i.e. if the pulsed pumping has repetition rate $ f= 8 \times
10^7 $~s$ {}^{-1} $, a photon pair is generated from one pump
pulse with probability $ 3.8 \times 10^{-4} $. Taking into account
the depth of the waveguide of the order of pump-field wavelength,
this probability is high. There are two reasons. The nonlinear
process exploits the largest element of the nonlinear tensor $ d $
of LiNbO$ {}_3 $ (this gives 2 orders of magnitude in comparison
with commonly used orientations). Also the down-converted fields
are confined in their transverse profiles into very narrow
regions, so the electric-field amplitude per one photon is high
which improves efficiency of the nonlinear process.

Signal-field intensity spectrum $ S_s $ determined according to
the formula
\begin{equation}   
 S_s(\omega_s) = \hbar\omega_s \int_{-\infty}^{\infty} d\omega_i
  |\Phi(\omega_s,\omega_i)|^2
\end{equation}
takes a gaussian form:
\begin{equation}   
 S_s(\omega_s) = s_s \exp\left[- \frac{(\omega_s - \omega_s^0 -
 \delta\omega_s^0)^2}{\sigma_{\omega_s}^2} \right] .
\label{23}
\end{equation}
Amplitude $ s_s $, width $ \sigma_{\omega_s} $, and shift $
\delta\omega_s^0 $ of the center for the signal-field intensity
spectrum assuming negligible coefficients $ {\cal G}_s $, $ {\cal
G}_i $, and $ {\cal G}_{si} $ are given as:
\begin{eqnarray}    
 s_s &=& |C_\Phi|^2 \exp(-2f_0) \frac{\sqrt{\pi}\hbar\omega_s^0\tau_p
  Z_p}{1+a_p^2} \nonumber \\
 & & \mbox{} \times \left[ \frac{\tau_p^2}{2(1+a_p^2)} +
  \frac{2}{\sigma_i^2} + \frac{Z_p^2 V_{pi}^2}{2} \right]^{-1/2} ,
  \\
 \sigma_{\omega_s} &=& \sqrt{ \frac{\tau_p^2}{2(1+a_p^2)} +
  \frac{2}{\sigma_i^2} + \frac{Z_p^2 V_{pi}^2}{2} } D_{f^r}^{-1/2},
  \label{25}
  \\
 \delta\omega_s^0 &=& 0;
\end{eqnarray}
$ D_{f^r} $ is given in Eq.~(\ref{20}). The general expressions
for parameters $ s_s $, $ \sigma_{\omega_s} $, and $
\delta\omega_s^0 $ can be found in Appendix A
[Eqs.~(\ref{ss}---\ref{dos})]. Characteristics of a gaussian
idler-field intensity spectrum $ S_{i}(\omega_i) $ can be derived
from symmetry. When frequency filters are not included, the
expression in Eq.~(\ref{25}) for width $ \sigma_{\omega_s} $ can
be further simplified:
\begin{equation}    
 \sigma_{\omega_s} = \frac{\sqrt{2}}{V_{si}} \sqrt{
  \frac{1}{Z_p^2} + \frac{ (1+a_p^2)V_{pi}^2}{\tau_p^2} } .
\label{27}
\end{equation}
This means that the signal-field spectral width $
\sigma_{\omega_s} $ (and similarly the idler-field spectral width
$ \sigma_{\omega_i} $) decreases with increasing pump-pulse
duration $ \tau_p $ and pump-beam transverse width $ Z_p $.
Assuming cw pumping, spectral widths $ \sigma_{\omega_s}^{\rm cw}
$ and $ \sigma_{\omega_i}^{\rm cw} $ are inversely proportional to
pump-beam transverse width $ Z_p $;
\begin{equation}    
 \sigma_{\omega_s}^{\rm cw} = \sigma_{\omega_i}^{\rm cw} =
 \frac{ \sqrt{2} }{V_{si} Z_p} .
\end{equation}
On the other hand the following expressions hold for the pump-beam
transverse width $ Z_p $ sufficiently large,
\begin{eqnarray}    
 \left. \sigma_{\omega_s}\right|_{Z_p \rightarrow \infty} &=&
  \frac{\sqrt{2}|V_{pi}|}{V_{si}} \frac{\sqrt{1+a_p^2}}{\tau_p},
  \nonumber \\
\left. \sigma_{\omega_i}\right|_{Z_p \rightarrow \infty} &=&
  \frac{\sqrt{2}|V_{ps}|}{V_{si}} \frac{\sqrt{1+a_p^2}}{\tau_p},
\end{eqnarray}
i.e. widths $ \sigma_{\omega_s} $ and $ \sigma_{\omega_i} $ are
inversely proportional to pump-pulse duration $ \tau_p $.
Dependence of the width $ \sigma_{\lambda_s} $ of the signal-field
intensity spectrum [$ \sigma_{\lambda_s} = 2\pi c / (\omega_s^0)^2
\sigma_{\omega_s} $] on pump-pulse duration $ \tau_p $ and
pump-beam transverse width $ Z_p $ is shown in Fig. \ref{fig2} for
the considered waveguide. Narrow spectra broad circa 1~nm are
generated if pump-pulse duration $ \tau_p $ is sufficiently long
and also pump-beam transverse width $ Z_p $ sufficiently wide. On
the other hand, intensity spectra wide several tens of nm can be
observed for ultrashort pump-pulses and strongly focused pump
beams. The reason for this behavior is that shortening of a pump
pulse and focusing of a pump beam lead to weakening of frequency
and phase matching conditions.

\begin{figure}    
 \resizebox{0.9\hsize}{!}{\includegraphics{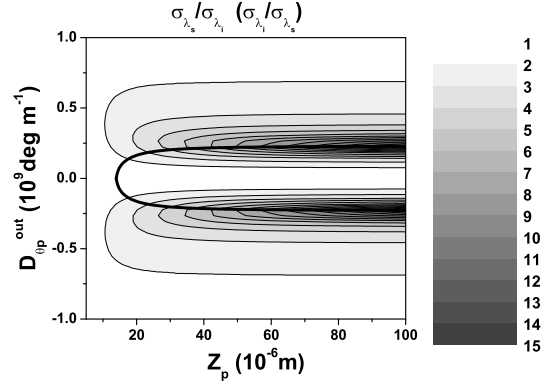}}
 \caption{Contour plot of ratio $ \sigma_{\lambda_s}/\sigma_{\lambda_i} $ of spectral widths
  of signal- and idler-field intensities as a function of parameter
  $ D_{\theta_p}^{\rm out} $ giving angular decomposition of pump-beam frequencies
  outside the waveguide and pump-beam transverse width $ Z_p $
  is shown. Ratio $ \sigma_{\lambda_i}/\sigma_{\lambda_s} $ is plotter for $ D_{\theta_p}^{\rm out} < 0 $.
  The bold curve indicating spectrally uncorrelated states is given using the formula
  in Eq.~(\ref{Dtps}) later. We have $ D_{\theta_p}^{\rm out} =
  \tilde{D}_{\theta_p}^{\rm out} (\omega_p^0)^2 / (2\pi c) $,
  $ \tilde{D}_{\theta_p}^{\rm out} = [n_0(\omega_p^0) \cos(\theta_p^0)] / \cos(\theta_p^{\rm out}) \tilde{D}_{\theta_p}
  + \sin(\theta_p^0) / \cos(\theta_p^{\rm out}) \left. dn_0(\omega_p)/d\omega_p
  \right|_{\omega_p=\omega_p^0} $ and
  $ \theta_p^{\rm out} = \arcsin[ n_0(\omega_p^0) \sin(\theta_p^0)] $; $ \tau_p =
  1 \times 10^{-13} $~s, values of the other parameters are given
  in Fig.~\ref{fig2}.  }
\label{fig3}
\end{figure}

Ratio $ \sigma_{\omega_s}/\sigma_{\omega_i} $ of spectral widths
of the signal- and idler-field intensities is important in some
applications \cite{URen2003}. For example, photon pairs used in
heralded single-photon sources should preferably be composed of
one photon with a narrow spectrum (convenient in propagation
through an optical fiber) and one photon with a wide spectrum
(leading to high detection efficiencies when post-selecting).
Provided that the pump beam is cross spectrally pure, the ratio $
\sigma_{\omega_s}/ \sigma_{\omega_i} $ is given by material
constants (group velocities of three fields) for given values of
pump-pulse duration $ \tau_p $ and pump-beam transverse width $
Z_p $. The ratio $ \sigma_{\omega_s}/\sigma_{\omega_i} $ equals 1
if the signal and idler fields are symmetric ($ \omega_s^0 =
\omega_i^0 $). On the other hand, parameter $ \tilde{D}_{\theta_p}
$ characterizing angular decomposition of pump-beam frequencies
can substantially change this ratio. Either narrowing or
broadening of an intensity spectrum of a down-converted field can
be reached by the change of values of parameter $
\tilde{D}_{\theta_p} $, as formulas in Eqs. (\ref{27}) and
(\ref{11}) indicate. Values around 10 for the ratio $
\sigma_{\omega_s}/ \sigma_{\omega_i} $ can be obtained for the
waveguide with values of parameters defined in Fig.~\ref{fig2}, as
documented in Fig.~\ref{fig3}. We note that high values of this
ratio occur when the two-photon spectral amplitude $ \Phi $ is
nearly spectrally uncorrelated (compare Fig.~\ref{fig8} later).

\section{Temporal properties of photon pairs}

Temporal properties of photon pairs can be conveniently described
using two-photon amplitude $ \Phi $ in the time domain that is
given as a Fourier transform of that in the frequency domain:
\begin{eqnarray}   
 \Phi(\tau_s,\tau_i) &=& \frac{1}{2\pi} \int_{-\infty}^{\infty}
  d\omega_s \int_{-\infty}^{\infty} d\omega_i
  \Phi(\omega_s,\omega_i) \nonumber \\
  & & \mbox{} \times \exp(-i\omega_s\tau_s)
  \exp(-i\omega_i\tau_i).
\label{30}
\end{eqnarray}
Because the two-photon spectral amplitude $ \Phi $ as given in
Eq.~(\ref{TPAO}) is gaussian, the two-photon amplitude $ \Phi $
defined in Eq.~(\ref{30}) takes also a gaussian form, that can be
found in Appendix A [Eq.~(\ref{TPAT})].

Photon flux $ N_s $ in the signal field is then determined along
the formula \cite{PerinaJr2006}
\begin{equation}   
 N_s(\tau_s) = \hbar \omega_s^0 \int_{-\infty}^{\infty} d\tau_i
 |\Phi(\tau_s,\tau_i)|^2 ,
\label{Ns}
\end{equation}
and attains the following gaussian form;
\begin{equation}   
 N_s(\tau_s) = n_s \exp\left[- \frac{(\tau_s -
 \delta\tau_s^0)^2}{\sigma_{\tau_s}^2} \right].
\label{Nss}
\end{equation}
Considering a pump pulse without chirp ($ a_p=0 $) and omitting
contributions in Eq.~(\ref{10}) given by coefficients $ {\cal G}_s
$, $ {\cal G}_i $, and $ {\cal G}_{si} $, we arrive at the
following simplified expressions:
\begin{eqnarray}    
 n_s &=& |C_\Phi|^2 \frac{\sqrt{\pi}\hbar\omega_s^0\tau_p
  Z_p}{ \sqrt{{\cal D}_f} } \nonumber \\
 & & \mbox{} \times \left[ \frac{\tau_p^2}{2} +
  \frac{2}{\sigma_s^2} + \frac{Z_p^2 V_{ps}^2}{2} \right]^{-1/2} ,
  \\
 \sigma_{\tau_s} &=& \sqrt{ \frac{\tau_p^2}{2} +
  \frac{2}{\sigma_s^2} + \frac{Z_p^2 V_{ps}^2}{2} } ,
 \label{sts} \\
 \delta\tau_s^0 &=& 0;
\end{eqnarray}
coefficient $ {\cal D}_f $ is given in Eq.~(\ref{Df}) in Appendix
A. The formula in Eq.~(\ref{sts}) shows that the signal-field
temporal width $ \sigma_{\tau_s} $ increases as the pump-pulse
duration $ \tau_p $ and pump-beam transverse width $ Z_p $
increase. Also the narrower the signal-field frequency filter, the
greater the temporal width $ \sigma_{\tau_s} $.

\begin{figure}    
 \resizebox{0.9\hsize}{!}{\includegraphics{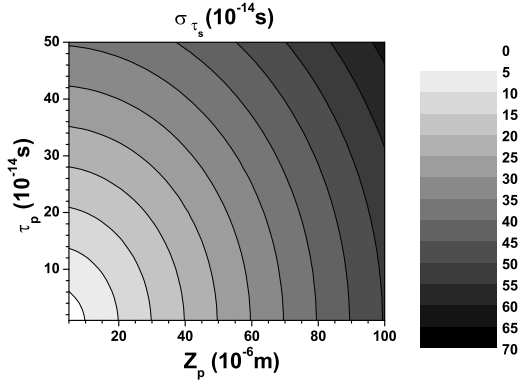}}
 \caption{Contour plot of temporal width $ \sigma_{\tau_s} $ of signal-field
  photon flux as a function of pump-pulse duration $ \tau_p $ and
  pump-beam transverse width $ Z_p $ is shown; values of the used parameters
  are given in Fig.~\ref{fig2}.}
\label{fig4}
\end{figure}

Depending on pump-pulse duration $\tau_p $ and pump-beam
transverse width $ Z_p $ width $ \sigma_{\tau_s} $ of the
signal-field intensity profile spreads from several tens to
several hundreds of fs (see Fig.~\ref{fig4}). Width $
\sigma_{\tau_s} $ of the signal-field intensity temporal profile
has to be greater than that characterizing the pump field ($
\tau_p/\sqrt{2} $) as the expression in Eq.~(\ref{sts}) confirms.
Also the greater the pump-beam transverse width $ Z_p $ the
greater the width $ \sigma_{\tau_s} $ as a consequence of
propagation of a signal photon along the waveguide. Angular
decomposition of pump-beam frequencies ($ D_{\theta_p} $) can
either broaden or shorten the pulsed photon fluxes $ N_s $ and $
N_i $ as follows from formulas in Eqs.~(\ref{sts}) and (\ref{11}).

Temporal and spectral widths of the down-converted fields are not
independent provided that the pump pulse is not chirped ($ a_p=0
$):
\begin{equation}  
 \frac{\sigma_{\omega_s}\sigma_{\tau_s} }{
  \sigma_{\omega_i}\sigma_{\tau_i}} = 1.
\end{equation}

Considering the symmetric case ($ \omega_s^0 = \omega_i^0 $),
cross spectrally pure pumping ($ D_{\theta_p} = 0 $) without chirp
($ a_p=0 $), and no frequency filters ($ \sigma_s, \sigma_i
\rightarrow \infty $), the following relation for the product of
temporal and spectral widths of the down-converted fields can be
derived:
\begin{equation}  
 \sigma_{\omega_a}\sigma_{\tau_a} = \frac{1}{2}
  \left( \frac{v_a\tau_p}{Z_p} + \frac{Z_p}{v_a \tau_p}
  \right) \ge 1, \hspace{3mm} a=s,i.
\label{37}
\end{equation}
Product $ \sigma_{\omega_a}\sigma_{\tau_a} $ written in
Eq.~(\ref{37}) goes to infinity for cw pumping.

Temporal properties as described by photon fluxes $ N_s $ and $
N_i $ are not accessible directly experimentally owing to very
short time scales. Thus interferometric setups are needed to
obtain these characteristics. Temporal overlap of the signal- and
idler-photon wave-functions can be detected in a Hong-Ou-Mandel
interferometer that provides a normalized coincidence-count rate $
R_{\rm n} $ as a function of mutual time delay $ \tau_l $ between
the signal and idler photons \cite{PerinaJr1999}:
\begin{equation}    
 R_{\rm n}(\tau_l) = 1 - \rho(\tau_l) ,
\end{equation}
where
\begin{eqnarray}   
 \rho(\tau_l) &=& \frac{1}{N} \int_{-\infty}^{\infty} dt_A \,
  \int_{-\infty}^{\infty} dt_B \, \nonumber \\
 & & \hspace{-7mm} {\rm Re} \left\{  \Phi(t_A,
  t_B- \tau_l) \Phi^*(t_B,t_A-\tau_l) \right\}
\label{39}
\end{eqnarray}
and the number $ N $ of photons generated in 1~s is given by
Eq.~(\ref{N}). Symbol $ {\rm Re} $ means the real part of an
argument. The shape of two-photon amplitude $ \Phi $ thus
determines the pattern of coincidence-count rate $ R_n(\tau_l) $
as discussed, e.g., in \cite{PerinaJr1999}.

Using the two-photon spectral amplitude $ \Phi $ written in
Eq.~(\ref{TPAO}) the normalized coincidence-count rate $ R_{\rm n}
$ can be derived in the form:
\begin{equation}   
 R_{\rm n}(\tau_l) = 1 - A \exp(-B\tau_l^2) \cos[(\omega_s^0-\omega_i^0)
  \tau_l] .
\label{Rn}
\end{equation}
Coefficients $ A $ and $ B $ are given as follows neglecting
coefficients $ {\cal G}_s $, $ {\cal G}_i $, and $ {\cal G}_{si} $
[the general expressions can be found in Eqs.~(\ref{A}) and
(\ref{B}) in Appendix A]:
\begin{eqnarray}   
 \left. A\right|_{\sigma_s,\sigma_i\rightarrow\infty} &=&
  \left[ 1+ \frac{ Z_p^2(1+a_p^2) }{4\tau_p^2} \frac{ (V_{ps}^2 -
  V_{pi}^2)^2 }{V_{si}^2} \right]^{-1/2} ,
  \label{41}
  \\
 B &=& \left[ \frac{2}{\sigma_s^2} + \frac{2}{\sigma_i^2} +
  \frac{Z_p^2 V_{si}^2}{2} \right]^{-1} .
  \label{42}
\end{eqnarray}

Visibility $ V $ of a coincidence-count pattern described by the
formula in Eq.~(\ref{Rn}) and defined along the expression ($
R_{\rm n,min} $ gives a minimum value of $ R_{\rm n} $)
\begin{equation}   
 V = \frac{ R_{\rm n}(\tau_l\rightarrow \infty) - R_{\rm n,min} }{
  R_{\rm n}(\tau_l\rightarrow \infty) + R_{\rm n,min} }
\end{equation}
is obtained in the form:
\begin{equation}  
 V = \frac{A}{2-A} .
\end{equation}
For cw pumping ($ \tau_p \rightarrow \infty $) and without
frequency filters, coefficient $ A \rightarrow 1 $ and also
visibility $ V \rightarrow 1 $. Visibility $ V $ equal to 1 is
also reached for the symmetric case ($ \omega_s^0 = \omega_i^0 $
leads to $ A=1 $).

On the other hand, coefficient $ B $ given in Eq.~(\ref{42})
determines width $ \Delta\tau_l $ of the coincidence-count dip
defined by the condition $ R_{\rm n}(\Delta\tau_l/2) = 1- A/2 $.
This relation can be rewritten into a transcendent equation,
\begin{equation}  
 \frac{1}{2} - \exp\left( \frac{B\Delta\tau_l^2}{4}\right)
  \cos\left( \frac{\omega_s^0-\omega_i^0}{2} \Delta\tau_l \right)
  = 0 ,
\end{equation}
that has solution in the range $ \Delta\tau_l \in (0,
2\pi/[\omega_s^0-\omega_i^0] ) $. We note that oscillations at
frequency $ \omega_s^0 - \omega_i^0 $ occur in the
coincidence-count pattern of rate $ R_{\rm n} $. Coefficient $ B $
depends only on pump-beam transverse width $ Z_p $ and widths $
\sigma_s $ and $ \sigma_i $ of frequency filters. The narrower the
frequency filters and the wider the pump-beam transverse width $
Z_p $, the smaller the value of coefficient $ B $ and also the
larger the width $ \Delta\tau_l $ of normalized coincidence-count
rate $ R_n $. We can see that width $ \Delta\tau_l $ of the
coincidence-count dip does not depend on pump-pulse duration $
\tau_p $ and pump-beam parameter $ D_{\theta_p} $. The reason is
that a Hong-Ou-Mandel interferometer monitors only the difference
of the signal- and idler-field frequencies $ 2\omega = \omega_s -
\omega_i $ that depends only on pump-beam transverse width $ Z_p $
in the considered configuration [compare Eq.~(\ref{TPAOo})].

\begin{figure}    
 \resizebox{0.8\hsize}{!}{\includegraphics{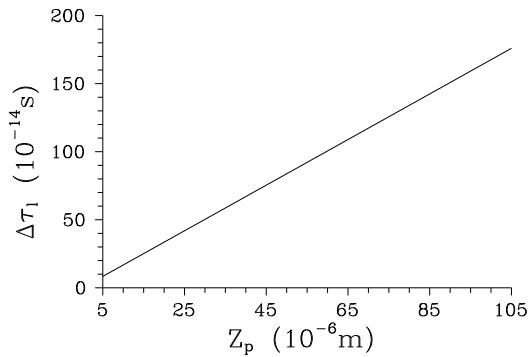}}

 \caption{Width $ \Delta\tau_l $ of the dip in normalized coincidence-count
  rate $ R_n $ in a Hong-Ou-Mandel interferometer
  as it depends on pump-beam transverse width $ Z_p $; values of
  the used parameters are written in Fig.~\ref{fig2}.}
\label{fig5}
\end{figure}

Width $ \Delta\tau_l $ of the coincidence-count dip can be
controlled changing pump-beam transverse width $ Z_p $ in a broad
range of its values, as documented in Fig.~\ref{fig5}. This
property may be useful in metrology applications (e.g., in
measurement of modal dispersion of a waveguiding structure).

\section{Entangled and separable states}

Entanglement between the signal- and idler-field frequencies can
be conveniently quantified using entropy of entanglement
\cite{Law2000,Law2004}. To determine entropy $ S_e $ of
entanglement we have to decompose a two-photon spectral amplitude
$ \Phi $ into Schmidt decomposition \cite{Parker2000}:
\begin{equation}   
 \Phi(\omega_s,\omega_i) = \sum_{n} \lambda_n \phi_{s,n}(\omega_s)
 \phi_{i,n}(\omega_i),
\label{SchmidtD}
\end{equation}
where $ \lambda_n $ are coefficients of the decomposition and
functions $ \phi_{s,n}(\omega_s) $ and $ \phi_{i,n}(\omega_i) $
form the Schmidt basis. This decomposition assuming a gaussian
two-photon spectral amplitude $ \Phi $ is found in Appendix B and
gives coefficients $ \lambda_n $ in the form of geometric
progression:
\begin{equation}  
 \lambda_n = \sqrt{1- \vartheta} \vartheta^{n/2} , \hspace{5mm}
  n=0,1, \ldots,\infty  .
\label{eigenval}
\end{equation}
Value of parameter $ \vartheta $ is derived from values of
parameters characterizing amplitude $ \Phi $ using Eqs.~(\ref{P})
and (\ref{vartheta}) in Appendix B. We note that $ \sum_n
\lambda_n^2 = 1 $ as a consequence of assumed normalization of the
amplitude $ \Phi $; $ \int d\omega_s \int d\omega_i
|\Phi(\omega_s,\omega_i)|^2 = 1 $.

Entanglement between the signal- and idler-field frequencies can
be determined using entropy $ S_e $ derived from eigenvalues $
\lambda_n $ of the Schmidt decomposition,
\begin{equation}   
 S_e = - \sum_n \lambda_n^2 \log_2(\lambda_n^2) ;
\label{Ee}
\end{equation}
symbol $ \ln_2 $ means logarithm of base 2. Using
Eq.~(\ref{eigenval}) we arrive at:
\begin{equation}   
 S_e = - \log_2(1-\vartheta) -
 \frac{\vartheta\log_2(\vartheta)}{1-\vartheta} .
\label{Se}
\end{equation}

As for possible values of entropy $ S_e $, there are two boundary
cases. If $ \vartheta \rightarrow 1 $ (also $ P \rightarrow 0 $
and $ |e_2| = e_{2c} $) then all eigenvalues $ \lambda_n $ are
equal, i.e. we have a maximally entangled state. On the other
hand, if $ \vartheta \rightarrow 0 $ (also $ P \rightarrow \infty
$ and $ e_{2c} \rightarrow 0 $) there is only one nonzero
eigenvalue $ \lambda_0 $; i.e. the two-photon spectral amplitude $
\Phi(\omega_s,\omega_i) $ factorizes and describes a separable
state useful, e.g., in linear quantum computation \cite{URen2003}.

\subsection{Entangled states}

Entangled states with high values of entropy $ S_e $ of
entanglement are generated if either pump-pulse duration $ \tau_p
$ is short or pump-beam transverse width $ Z_p $ is narrow, as the
analysis contained in Appendix B shows. The shorter the pump-pulse
duration $ \tau_p $ the greater the value of entropy $ S_e $. The
narrower the pump-beam transverse width $ Z_p $ the greater the
value of entropy $ S_e $. Also the wider the widths $ \sigma_s $
and $ \sigma_i $ of frequency filters the greater the values of
entropy $ S_e $.

On the other hand, widths $ \sigma_{\omega_s} $ and $
\sigma_{\omega_i} $ of the signal- and idler-field intensity
spectra increase with decreasing pump-pulse duration $ \tau_p $
and pump-beam transverse width $ Z_p $. The used frequency filters
make these spectra narrower. Comparing qualitatively this behavior
with that of entropy $ S_e $ of entanglement described above, we
can conclude that the wider the spectra of the signal and idler
fields, the better the entanglement of the signal and idler
fields.

Instead of characterizing entanglement by entropy $ S_e $ we can
judge it by a minimum number $ n_{\rm min} $ of eigenfunctions
(modes) from the Schmidt basis that restore a two-photon spectral
amplitude $ \Phi $ with probability $ p_{\rm min} $. The number $
n_{\rm min} $ satisfies the following inequalities:
\begin{equation}    
 \sum_{n=0}^{n_{\rm min}-1} \lambda_n^2 \le  p_{\rm min} \wedge
 \sum_{n=0}^{n_{\rm min}} \lambda_n^2 \ge  p_{\rm min} .
\label{nmin}
\end{equation}
The value of probability $ p_{\rm min} $ should be chosen with
respect to the precision of measurement.

\begin{figure}    
 {\raisebox{4.5 cm}{a)} \hspace{5mm}
 \resizebox{0.85\hsize}{!}{\includegraphics{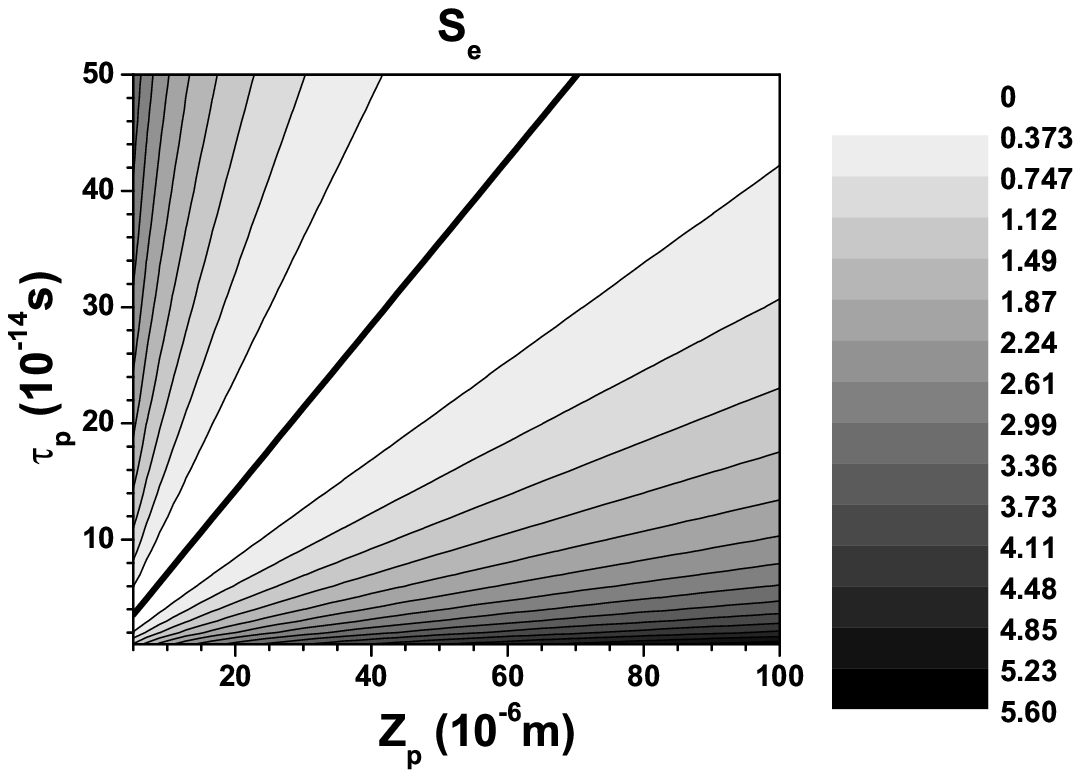}}

 \vspace{0mm}
 \raisebox{4.5 cm}{b)} \hspace{5mm}
 \resizebox{0.85\hsize}{!}{\includegraphics{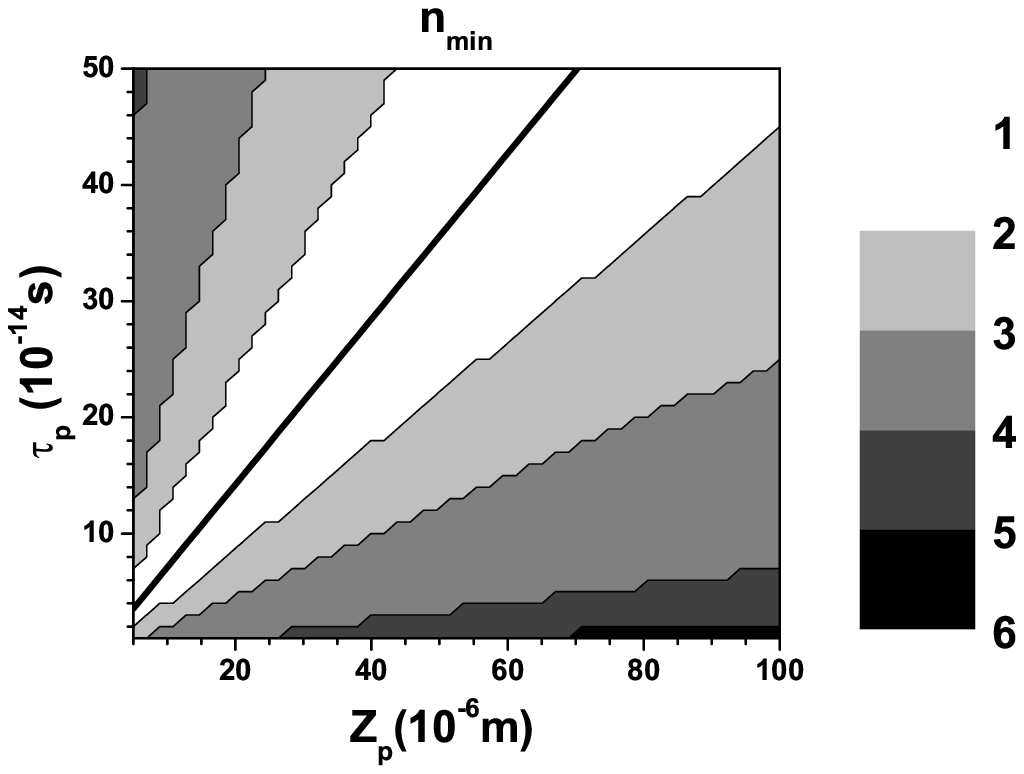}}}
 \vspace{0mm}

 \caption{Contour plots of entropy $ S_e $ of entanglement (a) and
  minimum number $ n_{\rm min} $ of modes in Schmidt
  decomposition ($ p_{\rm min} = 0.95 $) (b) as they depend on
  pump-pulse duration $ \tau_p $ and
  pump-beam transverse width $ Z_p $; bold curves in the middle of
  the plots are given by the condition in Eq.~(\ref{separ}) for
  spectrally uncorrelated states; values of
  the used parameters are given in Fig.~\ref{fig2}.}
\label{fig6}
\end{figure}

For the considered waveguide, entropy $ S_e $ of entanglement as
well as minimum number $ n_{\rm min} $ of modes are shown in
Fig.~\ref{fig6} as they depend on pump-pulse duration $ \tau_p $
and pump-beam transverse width $ Z_p $. We can see in
Fig.~\ref{fig6} that entanglement of the signal and idler fields
is rather weak in a broad area around the bold curve
characterizing spectrally uncorrelated (separable) states and so
we can approximate the generated state by a separable two-photon
spectral amplitude $ \Phi $. Entangled states for which several
modes in the Schmidt decomposition are necessary occur on the
borders of the contour plot, i.e. where the quantity $ \tau_p v_s
/ Z_p $ considerably differs from one [see Eq.~(\ref{separspec})
later]. Wide intensity spectra occur in this region (compare
Fig.~\ref{fig2}).

\begin{figure}    
 \resizebox{0.8\hsize}{!}{\includegraphics{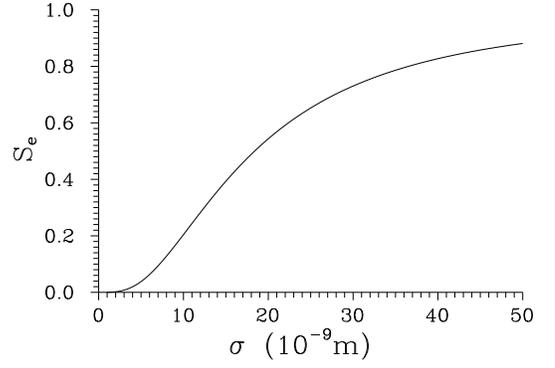}}

 \caption{Entropy $ S_e $ of entanglement as a function of
  width $ \sigma $ ($ \sigma_s = \sigma_i = \sigma $) of frequency
  filters in nm; $ \tau_p = 1\times 10^{-13} $~s, $ Z_p = 5 \times 10^{-6} $~m,
  $ p_{\rm min} = 0.95 $ and values of
  the other parameters are written in Fig.~\ref{fig2}.}
\label{fig7}
\end{figure}

The narrower the frequency filters the lower the values of entropy
$ S_e $ and thus the weaker the entanglement between the
down-converted fields, as demonstrated in Fig.~\ref{fig7}.
Assuming $ p_{\rm min} = 0.95 $ and values of the waveguide
parameters appropriate for Fig.~\ref{fig7}, filters narrower than
12~nm transform the down-converted fields into a separable state
whereas two independent modes are sufficient for wider filters.

\subsection{Spectrally uncorrelated (separable) states}

The condition for a separable state derived from the formula in
Eq.~(\ref{Se}), $ e_{2c} = 0 $, is fulfilled provided that $
f_{2si} = 0 $ as follows from the definition of coefficient $
e_{2c} $ in Eq.~(\ref{coefe}) in Appendix B. Separability of the
two-photon spectral amplitude $ \Phi $ written in Eq.~(\ref{TPAO})
is clearly visible in this case and shows that separable states
are generated only provided that the axes in which the quadratic
form given in Eq.~(\ref{TPAO}) is diagonal coincide with the $
\omega_s $ and $ \omega_i $ axes. Eigenvalues $ \mu_{1,2} $ of the
quadratic form written in the second line of Eq.~(\ref{TPAO}) are
given as follows:
\begin{equation}    
 \mu_{1,2} = \frac{ f_{2s} + f_{2i} \pm \sqrt{
 (f_{2s}-f_{2i})^2 + f_{2si}^2 } }{2} ,
\end{equation}
whereas angle $ \psi_{si} $ giving declination of the axes of the
diagonal quadratic form from the $ \omega_s $ and $ \omega_i $
axes is written as:
\begin{equation}   
 \tan(\psi_{si}) = \frac{ f_{2s} - f_{2i} \pm \sqrt{
 (f_{2s}-f_{2i})^2 + f_{2si}^2 } }{f_{2si}} .
\label{psisi}
\end{equation}
The limit $ f_{2si} \rightarrow 0 $ in Eq.~(\ref{psisi}) leads to
$ \psi_{si} = 0 $, i.e. the axes of diagonal quadratic form
coincide with the $ \omega_s $ and $ \omega_i $ axes.

Substituting expressions in Eqs.~(\ref{10}) and (\ref{11}) into
the separability condition $ f_{2si}=0 $ we arrive at:
\begin{eqnarray}   
 & & \mbox{} \hspace{-4mm} \frac{\tau_p^2}{1+ia_p} + Z_p^2 \left[
  \frac{\sin(\theta_p^0)}{v_p} + k_p^0\cos(\theta_p^0)
  \tilde{D}_{\theta_p} - \frac{1}{v_s} \right]  \nonumber \\
 & & \mbox{} \hspace{-4mm} \times \left[ \frac{\sin(\theta_p^0)}{v_p} + k_p^0\cos(\theta_p^0)
  \tilde{D}_{\theta_p} + \frac{1}{v_i} \right] + 2{\cal G}_{si} = 0 .
\label{separ}
\end{eqnarray}

Assuming fixed values for pump-pulse duration $ \tau_p $ and
pump-beam transverse width $ Z_p $ and no chirp ($ a_p = 0 $) the
condition in Eq.~(\ref{separ}) represents a quadratic equation for
parameter $ \tilde{D}_{\theta_p} $ of angular decomposition of
pump-beam frequencies and its solution takes the form:
\begin{eqnarray}   
 \left( \tilde{D}_{\theta_p} \right)_{1,2} &=& \frac{1}{2k_p^0\cos(\theta_p^0)} \left[ -
 \frac{2\sin(\theta_p^0)}{v_p} + \frac{1}{v_s} - \frac{1}{v_i}
 \right. \nonumber \\
 & & \left. \mbox{} \hspace{-2mm} \pm \sqrt{ \left( \frac{1}{v_s}
  + \frac{1}{v_i} \right)^2 - 4
  \frac{\tau_p^2}{Z_p^2} - 8 \frac{{\cal G}_{si} }{Z_p^2} } \right] .
\label{Dtp}
\end{eqnarray}
Solution for $ \tilde{D}_{\theta_p} $ written in Eq.~(\ref{Dtp})
exists only when argument of the square root in Eq.~(\ref{Dtp}) is
non-negative. Thus, there is a lower limit for possible values of
pump-beam transverse width $ Z_p $ keeping pump-pulse duration $
\tau_p $ fixed.

Considering the symmetric case ($ \omega_s^0 = \omega_i^0 $) and
neglecting coefficient $ {\cal G}_{si} $, the formula in
Eq.~(\ref{Dtp}) for $ \tilde{D}_{\theta_p} $ simplifies:
\begin{equation}     
 \left( \tilde{D}_{\theta_p} \right)_{1,2} = \pm \frac{1}{k_p^0} \sqrt{
  \frac{1}{v_s^2} - \frac{\tau_p^2}{Z_p^2} }
\label{Dtps}
\end{equation}
and is valid provided that $ Z_p \ge v_s \tau_p $. If the pump
beam is cross spectrally pure, i.e. $ \tilde{D}_{\theta_p} = 0 $,
the following condition assures the generation of a separable
state:
\begin{equation}     
 \frac{\tau_p}{Z_p} = \frac{1}{v_s} .
\label{separspec}
\end{equation}

\begin{figure}    
 \resizebox{0.8\hsize}{!}{\includegraphics{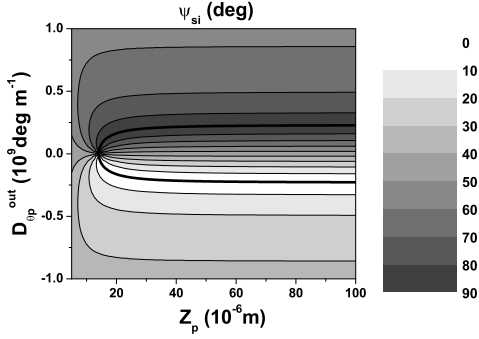}}

 \caption{Angle $ \psi_{si} $ giving orientation of axes of the diagonal quadratic form
  of two-photon spectral amplitude $ \Phi $ with respect to the $ \omega_s $
  and $ \omega_i $ axes as it depends on pump-beam transverse
  width $ Z_p $ and parameter $ D_{\theta_p}^{\rm out} $ giving angular decomposition
  of pump-beam frequencies outside the waveguide (for its determination, see Fig.~\ref{fig3}).
  The bold curve indicates spectrally uncorrelated states and is given by the formula
  in Eq.~(\ref{Dtps}); $ \tau_p = 1\times 10^{-13} $~s, values of
  the other parameters are the same as in Fig.~\ref{fig2}.}
\label{fig8}
\end{figure}

Changing values of parameters $ \tau_p $, $ Z_p $, and $
\tilde{D}_{\theta_p} $ photon pairs with an arbitrary gaussian
two-photon spectral amplitude $ \Phi $ can be generated
\cite{Walton2004}. An arbitrary orientation of axes of the
diagonal quadratic form of two-photon amplitude $ \Phi $ can be
reached changing the value of parameter $ \tilde{D}_{\theta_p} $
of angular decomposition of pump-beam frequencies, as demonstrated
in Fig.~\ref{fig8}. Parameters $ \tau_p $ and $ Z_p $ then control
spread of the two-photon spectral amplitude $ \Phi $ along the
axes given by angles $ \psi_{si} $ and $ \psi_{si} + \pi/2 $.
Comparison of graphs in Figs.~\ref{fig3} and \ref{fig8} with
special attention to frequency uncorrelated states leads to the
conclusion that the greater the pump-beam transverse width $ Z_p $
the greater the ratio $ \sigma_{\omega_s}/\sigma_{\omega_i} $ of
intensity spectral widths of the down-converted fields. Such
states have been found useful in quantum communication protocols.

\section{Experimental determination of entropy of entanglement}

If the pump field is not chirped ($ a_p =0 $), a generated photon
pair is described by a two-photon spectral amplitude $ \Phi $ in a
gaussian form written in Eq.~(\ref{TPAO}) that, apart from central
frequencies, is specified by three real parameters. These
parameters can be experimentally determined measuring widths of
the signal- ($ \sigma_{\omega_s} $) and idler-field ($
\sigma_{\omega_i} $) intensity spectra and width $ \Delta\tau_l $
of coincidence-count pattern in a Hong-Ou-Mandel interferometer,
as shown below.

Coefficient $ B $ describing width of the coincidence-count dip in
the Hong-Ou-Mandel interferometer can be obtained by fitting the
experimental curve $ R_{\rm n}(\tau_l) $ using the prescription
written in Eq.~(\ref{Rn}). The measured widths $ \sigma_{\omega_s}
$ and $ \sigma_{\omega_i} $ of signal- and idler-field intensity
spectra provide coefficient $ F $ [for definition, see
Eq.~(\ref{F}) in Appendix A]:
\begin{equation}  
 F = \frac{\sigma_{\omega_i}^2}{\sigma_{\omega_s}^2} .
\end{equation}
Further, Eqs.~(\ref{F}) and (\ref{B}) in Appendix A can be recast
into the form:
\begin{eqnarray}   
 f_{2i}^r &=& \frac{f_{2s}^r}{F} , \nonumber \\
 f_{2si}^r &=& \frac{F+1}{F} f_{2s}^r - \frac{1}{2B} .
\label{ff}
\end{eqnarray}
Substitution of Eqs.~(\ref{ff}) into Eqs.~(\ref{calD}) and
(\ref{sigmaos}) in Appendix A leads to a quadratic equation for
coefficient $ f_{2s}^r $:
\begin{equation}   
 - \frac{(F-1)^2}{F} (f_{2s}^{r})^2 + \left( \frac{F+1}{B} -
  \frac{2}{\sigma_{\omega_s}^2} \right) f_{2s}^r - \frac{F}{4B^2} =
  0 .
\label{quadr}
\end{equation}
Solution of Eq.~(\ref{quadr}) gives the value of coefficient $
f_{2s}^r $. For the symmetric case ($ \omega_s^0 = \omega_i^0 $),
$ F=1 $ and we have
\begin{equation}   
 f_{2s}^r =
  \frac{\sigma_{\omega_s}^2}{8B(\sigma_{\omega_s^2}-B)}.
\end{equation}
Values of coefficients $ f_{2i} $ and $ f_{2si} $ are then given
by formulas in Eqs.~(\ref{ff}). Coefficients $ f_{1s} $ and $
f_{1i} $ occurring in Eq.~(\ref{TPAO}) give only shifts on the
frequency axes $ \omega_s $ and $ \omega_i $ and can be neglected.
Also coefficient $ f_0 $ in Eq.~(\ref{TPAO}) giving normalization
can be omitted.

Knowing values of coefficients $ f_{2s} $, $ f_{2i} $, and $
f_{2si} $, entropy $ S_e $ of entanglement can finally be
determined along the formulas given in Eqs.~(\ref{coefe}),
(\ref{vartheta}), (\ref{P}), and (\ref{Se}).

\section{Conclusions}

Properties of the down-converted fields can be controlled in wide
ranges of values of characteristic parameters using pump-pulse
duration, pump-beam transverse width, and angular decomposition of
pump-beam frequencies in a waveguide with counter-propagating
down-converted fields and transverse pumping. Widths of intensity
spectra of the down-converted fields may vary from nanometers to
tens of nanometers. Durations of the down-converted pulsed fields
extend from tens of fs to several ps. Both entangled and separable
(spectrally uncorrelated) photon pairs useful in linear quantum
computation can be generated. Also attainable widths of a
coincidence-count dip in a Hong-Ou-Mandel interferometer lie in a
broad interval that is of interest in metrology applications.
Using the measured spectral widths of the signal and idler fields
and width of the coincidence-count dip, entropy of entanglement as
well as parameters characterizing a two-photon spectral amplitude
can be obtained.

The considered nonlinear waveguide is promising as a versatile
source of photon pairs in near future provided that efficiency of
the nonlinear process is increased.

\acknowledgments The author thanks V. Pe\v{r}inov\'{a} and A.
Luk\v{s} for their help with calculations and M. Centini for
useful discussions. Support by projects IAA 100100713 of GA AS CR,
1M06002, COST OC P11.003, and AVOZ 10100522 of the Czech Ministry
of Education is acknowledged.

\appendix

\section{General formulas for physical quantities characterizing
the emitted photon pairs}

\subsection{Spectral properties}

Substitution of the general form of two-photon spectral amplitude
$ \Phi $ as given in Eq.~(\ref{TPAO}) into Eq.~(\ref{N}) results
in the following expression for the number $ N $ of generated
photon pairs in 1~s:
\begin{equation}   
 N = |C_\Phi|^2 \exp(-2f_0) \frac{\pi Z_p\tau_p}{(1+a_p^2)\sqrt{{\cal D}_{f^r}}}
  {\cal E}_{f^r} ,
\label{A1}
\end{equation}
where
\begin{equation}   
 {\cal D}_{f^r} = 4f_{2s}^r f_{2i}^r - (f_{2si}^r)^2
\label{calD}
\end{equation}
and
\begin{eqnarray} 
 {\cal E}_{f^r} &=& \exp\left( 2\frac{f_{2s}^rf_{1i}^2 +
  f_{2i}^rf_{1s}^2 - f_{2si}^rf_{1s}f_{1i} }{ {\cal D}_{f^r}}
  \right) . \nonumber \\
  & &
\label{calE}
\end{eqnarray}
Superscript $ r $ indicates the real part of a given complex
coefficient.

A gaussian signal-field intensity spectrum $ S_s(\omega_s) $ as
written in Eq.~(\ref{23}) is determined by its amplitude $ s_s $,
width $ \sigma_{\omega_s} $, and shift $ \delta\omega_s^0 $ of the
center that are given in general by the following expressions:
\begin{eqnarray}      
 s_s &=& |C_\Phi|^2 \exp(-2f_0) \frac{\sqrt{\pi}\hbar\omega_s\tau_p
  Z_p}{ \sqrt{2}(1+a_p^2)} \frac{1}{\sqrt{f_{2i}^r}}
  {\cal E}_{f^r} ,
  \label{ss} \\
 \sigma_{\omega_s} &=& \sqrt{\frac{2f_{2i}^r}{{\cal D}_{f^r}}} ,
 \label{sigmaos}
  \\
 \delta\omega_s^0 &=& - \frac{2f_{2i}^r f_{1s} - f_{2si}^r f_{1i} }{
  {\cal D}_{f^r}},
  \label{dos}
\end{eqnarray}
where coefficients $ {\cal D}_{f^r} $ and $ {\cal E}_{f^r} $ are
given in Eqs. (\ref{calD}) and (\ref{calE}). Expressions for
amplitude $ s_i $, width $ \sigma_{\omega_i} $, and shift $
\delta\omega_i^0 $ belonging to the idler-field intensity spectrum
$ S_i(\omega_i) $ can be obtained from those written in
Eqs.~(\ref{ss}---\ref{dos}) by an exchange of indices $ s $ and $
i $.

Relations among parameters of the signal- and idler-field
intensity spectra can be established for a two-photon spectral
amplitude $ \Phi $ written in Eq.~(\ref{TPAO}):
\begin{eqnarray}    
 s_i &=& \frac{\omega_i^0}{\omega_s^0}
  \sqrt{F} s_s , \\
 \sigma_{\omega_i} &=& \frac{1}{\sqrt{F}} \sigma_{\omega_s} .
 \label{sigmaoi}
\end{eqnarray}
Symbol $ F $,
\begin{equation}   
 F = \frac{f_{2s}^r}{f_{2i}^r} ,
\label{F}
\end{equation}
gives the ratio of parameters that characterize a gaussian
two-photon spectral amplitude $ \Phi $.

\subsection{Temporal properties}

Substituting the expression in Eq.~(\ref{TPAO}) for two-photon
spectral amplitude $ \Phi(\omega_s,\omega_i) $ into the definition
of amplitude $ \Phi(\tau_s,\tau_i) $ occurring in Eq.~(\ref{30}),
we arrive at
\begin{eqnarray}   
 \Phi(\tau_s,\tau_i) &=& C_\phi \exp(-f_0) \sqrt{\frac{Z_p\tau_p}{(1+a_p^2){\cal
  D}_f}} \nonumber \\
  & & \hspace{-10mm} \mbox{} \times \exp\left[- \frac{
  f_{2i}(\tau_s-if_{1s})^2 + f_{2s}(\tau_i-if_{1i})^2 }{ {\cal D}_f } \right] \nonumber \\
  & & \hspace{-10mm} \mbox{} \times \exp\left[ \frac{ f_{2si}(\tau_s-if_{1s})
  (\tau_i-if_{1i}) }{ {\cal D}_f } \right]
\label{TPAT}
\end{eqnarray}
and
\begin{equation}  
 {\cal D}_f = 4 f_{2s} f_{2i} - f_{2si}^2 .
\label{Df}
\end{equation}

Parameters of the gaussian form of signal-field photon flux $ N_s
$ written in Eq.~(\ref{Nss}) [amplitude $ n_s $, width $
\sigma_{\tau_s} $, and shift $ \delta\tau_s^0 $] are given for the
general form of two-photon spectral amplitude $ \Phi $ in
Eq.~(\ref{TPAO}) by the following formulas, similarly as in the
case of intensity spectra:
\begin{eqnarray}      
 n_s &=& |C_\Phi|^2 \exp(-2t_0) \frac{\sqrt{\pi}\hbar\omega_s^0\tau_p
  Z_p}{ \sqrt{2}(1+a_p^2)} \frac{1}{|{\cal D}_f|} \frac{1}{\sqrt{t_{2i}}}
  {\cal E}_{t} , \nonumber \\
  & &
  \label{ns} \\
 \sigma_{\tau_s} &=& \sqrt{\frac{2t_{2i}}{{\cal D}_t}} ,
  \\
 \delta\tau_s^0 &=& - \frac{2t_{2i} t_{1s} - t_{2si} t_{1i} }{
  {\cal D}_t} ,
  \label{dts}
\end{eqnarray}
and
\begin{eqnarray}  
 {\cal D}_t &=& 4t_{2s} t_{2i} - t_{2si}^2
\label{calDt} , \\
 {\cal E}_t &=& \exp\left( 2\frac{t_{2s} t_{1i}^2 +
  t_{2i} t_{1s}^2 - t_{2si} t_{1s}t_{1i} }{ {\cal D}_t}
  \right) .
\label{Et}
\end{eqnarray}
Coefficients $ t $ occurring in the above Eqs.~(\ref{ns}
---\ref{Et}) are given as:
\begin{eqnarray}   
 t_{2s} &=& {\rm Re} \{f_{2i}/{\cal D}_f\} , \nonumber \\
 t_{2i} &=& {\rm Re} \{f_{2s}/{\cal D}_f\} , \nonumber \\
 t_{2si} &=& - {\rm Re} \{f_{2si}/{\cal D}_f\} , \nonumber \\
 t_{1s} &=& {\rm Im} \{(2f_{2i}f_{1s} - f_{2si}f_{1i})/{\cal D}_f \}
  ,\nonumber \\
 t_{1i} &=& {\rm Im} \{(2f_{2s}f_{1i} - f_{2si}f_{1s})/{\cal D}_f \}
  ,\nonumber \\
 t_0 &=& f_0 - {\rm Re} \{ (f_{2i}f_{1s}^2 + f_{2s}f_{1i}^2 -
  f_{2si}f_{1s}f_{1i})/{\cal D}_f \} ; \nonumber \\
  & &
\end{eqnarray}
$ {\rm Re} $ ($ {\rm Im} $) denotes a real (imaginary) part of an
expression. Similarly as in the case of spectral properties,
relations among parameters of the signal and idler fields can be
derived:
\begin{eqnarray}    
 n_i &=& \frac{\omega_i^0}{\omega_s^0}
  \sqrt{T} n_s , \\
 \sigma_{\tau_i} &=& \frac{1}{\sqrt{T}} \sigma_{\tau_s}.
\end{eqnarray}
Parameter $ T $,
\begin{equation}   
 T = \frac{t_{2i}}{t_{2s}},
\end{equation}
gives the ratio of parameters that characterize a gaussian
two-photon amplitude $ \Phi $ in the time domain. If the pulsed
pump field is not chirped, we have $ T = 1/F $.

Coefficients $ A $ and $ B $ characterizing a coincidence-count
pattern in a Hong-Ou-Mandel interferometer as given in
Eq.~(\ref{Rn}) are determined for the general form of two-photon
spectral amplitude $ \Phi $ as follows:
\begin{eqnarray}   
 A &=& \sqrt{\frac{ {\cal D}_{f^r} }{ (f_{2s} + f_{2i}^*)^2 -
  (f_{2si}^{r})^2 }} \exp \left[ \frac{ (f_{1s}+ f_{1i})^2 }{
  2(f_{2s} + f_{2i}^* + f_{2si}) } \right] \nonumber \\
 & & \mbox{} \times {\cal E}_{f^r}^{-1} ,
 \label{A}
  \\
 B &=& \frac{1}{ 2(f_{2s} + f_{2i}^* - f_{2si}) } ,
 \label{B}
\end{eqnarray}
and coefficients $ {\cal D}_{f^r} $ and $ {\cal E}_{f^r} $ are
given by formulas in Eqs.~(\ref{calD}) and (\ref{calE}).

\section{Schmidt decomposition of two-photon spectral amplitude $ \Phi $}

In order to determine Schmidt decomposition of two-photon spectral
amplitude $ \Phi $ as given in Eq.~(\ref{SchmidtD}) we need
reduced statistical operators belonging to the signal and idler
fields. Statistical operator $ \hat{\rho_s} $ of the signal field
can be written in the form:
\begin{equation}   
 \hat{\rho}_s = \int_{-\infty}^{\infty} d\omega'_s
  \int_{-\infty}^{\infty} d\omega_s \Psi_s(\omega'_s,\omega_s)
   \hat{a}_s^\dagger(\omega'_s) |{\rm vac} \rangle
   \langle {\rm vac}| \hat{a}_s(\omega_s).
\end{equation}
Weighting function $ \Psi_s $ determined along the formula
\begin{equation}   
 \Psi_s(\omega'_s,\omega_s) = \int_{-\infty}^{\infty} d\omega_i
  \Phi(\omega'_s,\omega_i) \Phi^*(\omega_s,\omega_i)
\end{equation}
takes the following form using Eq.~(\ref{TPAO}) for the two-photon
spectral amplitude $ \Phi $:
\begin{eqnarray}   
 \Psi_s(\omega_s',\omega_s) &=& |\tilde{C}_\Psi|^2 \exp\left(-e_2
  \omega_s^2 - e_2^* \omega'_s{}^{2} + 2e_{2c}\omega_s\omega'_s \right)
  \nonumber \\
 & & \mbox{} \times \exp\left(
  e_1 \omega_s+ e_1^* \omega'_s \right),
\label{Psis}
\end{eqnarray}
and
\begin{eqnarray}   
 e_2 &=& f_{2s} - \frac{f_{2si}^2}{8f_{2i}^r} , \nonumber \\
 e_{2c} &=& \frac{|f_{2si}|^2}{8f_{2i}^r} , \nonumber \\
 e_1 &=& f_{1s} - \frac{f_{1i}f_{2si}}{(f_{2i}^{r})^2}.
\label{coefe}
\end{eqnarray}
Normalization constant $ \tilde{C}_\Psi $ introduced in
Eq.~(\ref{Psis}),
\begin{equation}  
 |\tilde{C}_\Psi|^2 = \sqrt{\frac{ 2(e_2^r - e_{2c})}{\pi}}
  \exp\left[ - \frac{(e_1^{r})^2}{2(e_2^r-e_{2c})} \right] ,
\end{equation}
guarantees normalization of the signal-field statistical operator
$ \hat{\rho}_s $ such that $ \int_{-\infty}^{\infty} d\omega_s
\Psi (\omega_s,\omega_s) = 1 $. Statistical operator $
\hat{\rho}_i $ of the idler field can be expressed similarly as
that for the signal field.

Coefficients $ \lambda_n $ and functions $ \phi_{s,n} $ and $
\phi_{i,n} $ occurring in the Schmidt decomposition in
Eq.~(\ref{SchmidtD}) are given as solutions of the following
integral equations:
\begin{equation}  
 \int_{-\infty}^{\infty} d\omega_a \Psi_a(\omega'_a,\omega_a)
  \phi_{a,n}(\omega_a) = \lambda_n^2 \phi_{a,n}(\omega'_a),
  \hspace{5mm} a=s,i.
\label{inteq}
\end{equation}

Using linear substitution the kernel $ \Psi_s $ in
Eq.~(\ref{Psis}) can be transformed into the form:
\begin{equation}    
 \Psi(x,y) = \exp[ -(1+P)(x^2 + y^2) + 2xy ]
\label{Psi}
\end{equation}
and
\begin{equation}   
 P = \frac{|e_2|}{e_{2c}} - 1 .
\label{P}
\end{equation}

It can be shown \cite{Perinova2006} that the following functions $
\phi_n $ obey the integral equation in Eq.~(\ref{inteq}) for
kernel $ \Psi $ defined in Eq.~(\ref{Psi}):
\begin{eqnarray}   
 \phi_n(x) &=& \sqrt{ \frac{ \sqrt{ 1- \vartheta^2} }{2^nn! \sqrt{
  \pi\vartheta} } } \exp \left( - \frac{1-\vartheta^2}{2\vartheta}
  x^2 \right) \nonumber \\
 & & \hspace{-15mm} \mbox{} \times  H_n\left( \sqrt{ \frac{1-\vartheta^2}{\vartheta} } x
  \right) , \hspace{5mm} n=0,1,\ldots, \infty ;
\end{eqnarray}
symbols $ H_n $ denote Hermite polynomials. The corresponding
eigenvalues $ \lambda_n^2 $ form a geometric progression:
\begin{equation}   
 \lambda_n^2 = \sqrt{\pi\vartheta} \vartheta^n , \hspace{5mm}
  n=0,1,\ldots,\infty
\end{equation}
and parameter $ \vartheta $ is given as follows:
\begin{equation}  
 \vartheta = 1 + P - \sqrt{P^2+2P} .
\label{vartheta}
\end{equation}

A maximally entangled state according to entropy $ S_e $ of
entanglement determined in Eq.~(\ref{Se}) is reached if all
eigenvalues $ \lambda_n $ are equal, i.e. when $ P \rightarrow 0
$. To understand this condition, we express coefficient $ P $
defined in Eq.~(\ref{P}) in terms of coefficients $ f $:
\begin{eqnarray}   
 P = \sqrt{ 1+ \frac{16f_{2i}^r}{|f_{2si}|^4} \left( 4|f_{2s}|^2
  f_{2i}^r - {\rm Re} \{f_{2s} f_{2si}^{*2} \} \right) } -1 .
\label{PP}
\end{eqnarray}
It can be shown that coefficient $ P $ goes to zero if coefficient
$ {\cal D}_f $ defined in Eq.~(\ref{Df}) goes to zero too.
Provided that coefficients $ {\cal G}_s $, $ {\cal G}_i $, and $
{\cal G}_{si} $ can be omitted, the condition $ {\cal D}_f = 0 $
is fulfilled only in the limit $ \tau_p \rightarrow 0 $ and $ Z_p
\rightarrow 0 $. For this reason, we investigate the behavior of
coefficient $ {\cal D}_f $ with respect to pump-pulse duration $
\tau_p $, pump-beam transverse width $ Z_p $, and widths $
\sigma_s $ and $ \sigma_i $ of frequency filters in the area
around $ \tau_p = 0 $ or $ Z_p = 0 $.

For fixed values of $ Z_p $, $ \sigma_s $, and $ \sigma_i $ and
around $ \tau_p = 0 $ the following equality holds:
\begin{eqnarray}   
 \left. \frac{ \partial {\cal D}_f}{\partial(\tau_p^2)} \right|_{Z_p,
  \sigma_s, \sigma_i} &=& \frac{1}{1+ia_p}
  \left[ \frac{V_{si}^2 Z_p^2}{4} + \frac{1}{\sigma_s^2} + \frac{1}{\sigma_i^2}
  \right. \nonumber \\
 & & \left. \mbox{}
  + {\cal G}_s + {\cal G}_i - {\cal G}_{si} \right] .
\label{Ddtp}
\end{eqnarray}
Also fixing values of $ \tau_p $, $ \sigma_s $, and $ \sigma_i $
and being around $ Z_p = 0 $ we arrive at:
\begin{eqnarray}   
 \left. \frac{ \partial {\cal D}_f}{\partial (Z_p^2)} \right|_{\tau_p,
  \sigma_s, \sigma_i} &=& \frac{1}{1+ia_p}
 \frac{V_{si}^2 \tau_p^2}{4} + V_{ps}^2 \left( \frac{1}{\sigma_i^2}
  + {\cal G}_i \right)  \nonumber \\
 & & \hspace{-6mm} \mbox{} + V_{pi}^2 \left( \frac{1}{\sigma_s^2}
  + {\cal G}_s \right) - V_{ps}V_{pi}{\cal G}_{si}.
\label{DdZp}
\end{eqnarray}
If $ ( \left. dn_0(\omega_a)/d\omega_a
\right|_{\omega_a=\omega_a^0} ) \omega_a^0 \ll n_a $ for $ a = s,i
$ then $ g_{2si} = 2 \sqrt{g_{2s}g_{2i}} $ and the derivatives in
Eqs.~(\ref{Ddtp}) and (\ref{DdZp}) are positive ($ a_p = 0 $ is
assumed). This means that if the value of pump-pulse duration $
\tau_p $ is small [lower than that given by the condition in
Eq.~(\ref{separ})] the shorter the pump-pulse duration $ \tau_p $
the more the down-converted fields are entangled. Similarly it
holds for sufficiently small values of pump-beam transverse width
$ Z_p $ that the narrower the pump-beam transverse width $ Z_p $
the more the down-converted fields are entangled.

We get the following expressions for derivatives of coefficient $
{\cal D}_f $ with respect to widths $ \sigma_s $ and $ \sigma_i $
of frequency filters:
\begin{eqnarray}    
 \left. \frac{\partial{\cal D}_f}{\partial (\sigma_s^2)}
 \right|_{\tau_p, Z_p} &=& - \frac{1}{\sigma_s^4} \left[
  \frac{\tau_p^2}{1+ia_p} + V_{pi}^2 Z_p^2
  + \frac{1}{\sigma_{i}^2} + {\cal G}_i \right] , \nonumber \\
 \left. \frac{\partial{\cal D}_f}{\partial (\sigma_i^2)}
 \right|_{\tau_p, Z_p} &=& - \frac{1}{\sigma_i^4} \left[
  \frac{\tau_p^2}{1+ia_p} + V_{ps}^2 Z_p^2
  + \frac{1}{\sigma_{s}^2} + {\cal G}_s \right] . \nonumber \\
 & &
\label{Dds}
\end{eqnarray}
For our waveguide, $ {\cal G}_s > 0 $ and $ {\cal G}_i > 0 $ and
this means that the derivatives of coefficient $ {\cal D}_f $
written in Eqs.~(\ref{Dds}) are negative ($ a_p = 0 $ is assumed).
Thus the wider the frequency filters, the smaller the value of
coefficient $ {\cal D}_f $ and the more the down-converted fields
are entangled.

\end{document}